\documentstyle[11pt,newpasp,twoside,psfig]{article}
\def\edcomment#1{\iffalse\marginpar{\raggedright\sl#1\/}\else\relax\fi}
\reversemarginpar

\begin{document}
\title{
Formation of the optical spectra of the coolest M-
and L-dwarfs and lithium abundances in their atmospheres
}
\author{Yakiv V. Pavlenko}
\affil{
 Main Astronomical Observatory of NAS, Golosiiv woods, 03680, Kyiv-127,
 Ukraine}

\begin{abstract}
Theoretical aspects of modeling of spectra of late M- and
L-dwarfs are discussed. We show, that the processes of
formation of spectra of M- and L-dwarfs are basically different.
Instead of the case of M-dwarfs, atoms of Ti and V
should be depleted into grains in the atmospheres of L-dwarfs.
Overall shape of
the L-dwarf spectra is governed by the K I + Na I resonance line
wings of the huge strength. To fit lithium lines observed in
spectra of the coolest dwarfs we used two additional suggestions:
a) there are some {\em extra} depletions of molecular species
absorbed in the optical spectra of L-dwarfs; b) there may
be (a few?) additional (``dusty''?) opacity sources in their
atmospheres. Problems of lithium line formation and the
``natural'' limitation of their use for the ``lithium test'' for
the case of L-dwarfs are considered.

\end{abstract}

\section{Introduction}

A few definitions of the ``unconventional''
spectral classification of low mass stars
and substellar objects are used in this paper:   \\

{\bf M-dwarfs~} are objects (stars + young brown dwarfs) with
4000 $> T_{\rm eff} >$ 2200 K. Their spectra are governed by
molecular bands of TiO and VO (in the optical part of the
spectrum) , H$_2$O (in the red).  \\

{\bf L-dwarfs~} are objects (brown dwarfs + stars) with 1000 $<$
$T_{\rm eff}$ $<$  2200 K (cf. Martin et al. 1997).
 Optical spectra of L-dwarfs are governed by the K I + Na I
lines and molecular bands of CrH + FeH (in the optical spectrum),
H$_2$O (in the red).  \\

{\bf T-dwarfs~} are super-giant, planet-like
objects (big Jupiters?)
with T$_{\rm eff} <$ 1000 K. Their spectra are formed
by the "dusty opacities" and methane + H$_2$O + ...? absorption
(Strauss et al. 1999, Burgasser et al. 1999)\\

For the time being most of the known  brown dwarfs are actually 
recognized by the detection of  the  Li\,{\sc i}  resonance 
doublet  in  their  spectra   (Rebolo et al. 1996, Mart\'\i n et 
al. 1997a,  Kirkpatrick  et  al. 1999). Indeed, temperatures in 
the interiors of brown dwarfs are not high enough to burn lithium 
(Rebolo et al. 1992).

\section{Procedure}

The computations of synthetical spectra of M- and L-dwarfs are
carried out by program WITA5, which is a modified version of the
program WITA31 used by Pavlenko (1997). The
modifications  were  aimed  to incorporate ``dusty  effects''
that  affect  the  chemical equilibrium  and  radiative
transfer  processes  in  very    cool atmospheres.

We have used the set  of  Tsuji's  (1999)
``dusty'' (C-type)  LTE  model  atmospheres.  These  models  were
computed for the case of segregation phase of dust and gas.

Chemical equilibrium was computed for  the  mix  of  $\approx$100
molecular species.
To take into account the effect of the oversaturation,  we
reduced  the abundances of those molecular species  down  to  the
equilibrium values (Pavlenko 1998).

In L-dwarf atmospheres the additional opacity (AdO) could appear due
to molecular and/or dust absorption and/or scattering. We  have
modelled  the additional  opacity with  a simple law of the form
$a_{\circ} \ (\nu / \nu_{\circ})^N$,  with $N$\,= 1 -\,4 (see
Pavlenko, Zapatero Osorio \& Rebolo 2000 for more details).

\section{M-dwarf spectra}

Lithium lines observed in spectra of the late M-dwarfs are well
known tracers of their evolution.
Completely convective M-dwarfs are
very effective lithium destroyers, therefore cool pre-main
sequence (PMS) stars are expected to preserve their initial
lithium only during their first few million's years (Fig.3, see
also Magazzu, Rebolo \& Pavlenko 1992,
Oppenheimer et al. 1997, Pavlenko (1997, 1997a),
Pavlenko \& Oppenheimer 1998).

Lithium lines in spectra of M-dwarfs are formed at the background
of mighty TiO bands (Fig.1). Only cores of the saturated Li
lines may be observed in the real spectra (Pavlenko et al. 1995).
To estimate of the abundance of lithium one may use
``pseudoequivalent widths'' of lithium lines, i.e. $W_{\lambda}$
measured in
respect to the local pseudocontinuum formed by molecular bands
around Li lines (see also Pavlenko 1997a).

Sure, the better way of the quantitatively determination
log N(Li) is the use of synthetical spectra (Fig. 2).

\begin{figure}
\psfig{file=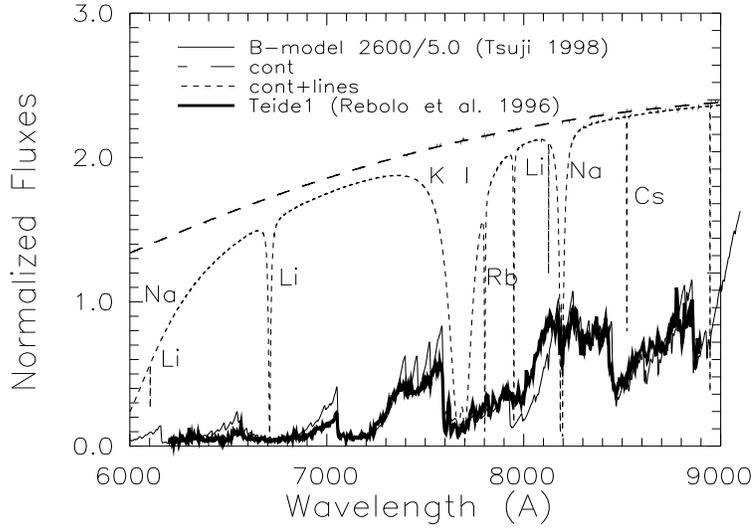,height=7cm,width=10cm}
\caption[]{
 Comparison of the observed
spectrum of the Pleiade's young brown dwarf Teide1 (Rebolo et al. 1996) 
with theoretical
spectra computed
using C-model atmosphere ($T_{\rm eff}$=2600~K, log g =5.0) of Tsuji (1999).
The strongest atomic lines which may be observed
in M- and L- spectra and the real (theoretical) continuum are shown.}
\end{figure}

\begin{figure}
\psfig{file=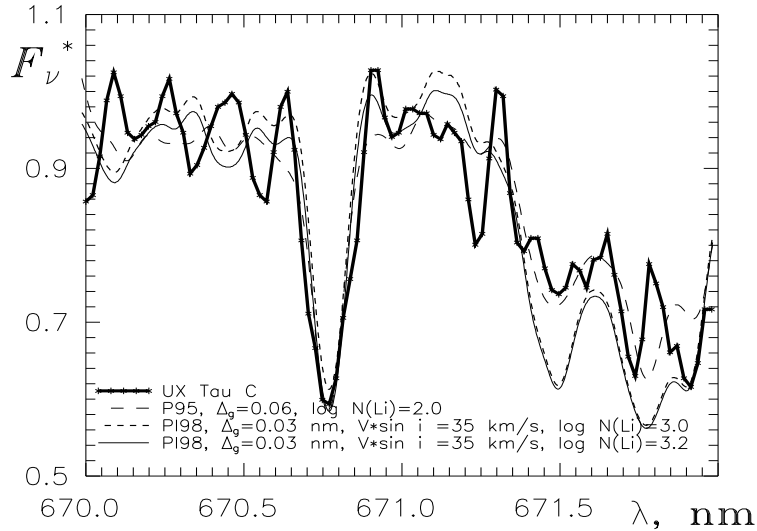,height=7cm,width=10cm}
\caption[]{Fit to the observed Li I resonance doublet
in UX Tau C spectrum (Magazzu et al. 1991) with the
list of TiO lines of Plez (1998) and
model atmosphere 3100/4.5 (Allard \& Hauschildt 1995)}
\end{figure}

\section{L-dwarf spectra}

Due to depletion of the Ti and VO into grains a structure of the
optical spectra of L-dwarfs becomes more simple in comparison with
M-dwarfs. The overall spectral energy distribution (SED)
is governed by absorption of
resonance doublets of K I and Na I which have pressure broadened
wings extended up to thousands \AA (Fig.3). Furthermore, Pavlenko et al.
(2000) showed that L-dwarf optical spectra are affected by
the additional
(``dusty'') absorption and/or scattering.

\begin{figure}
\psfig{file=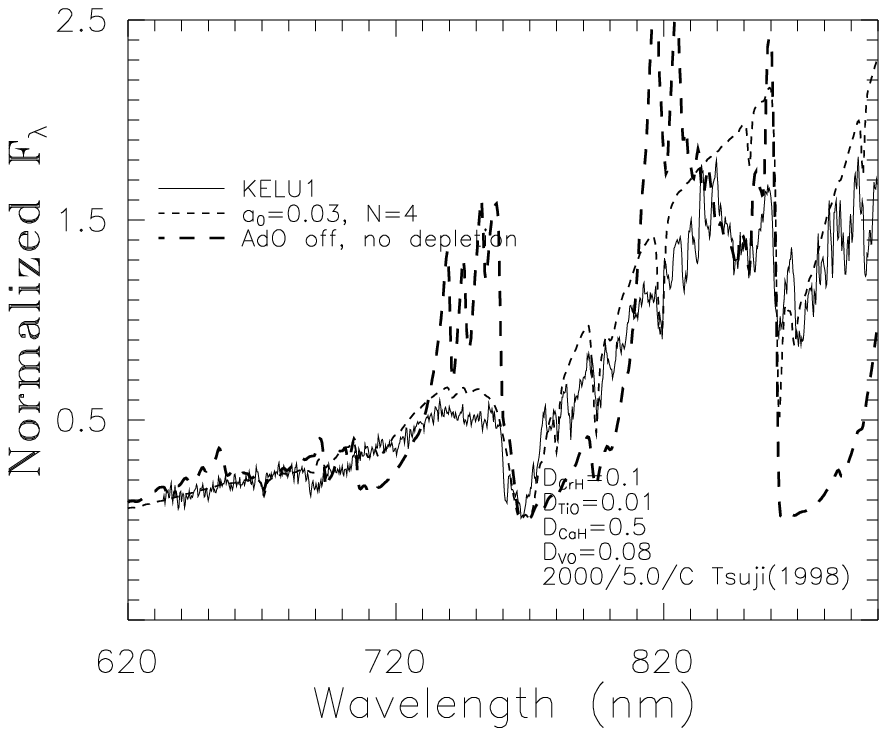,height=7cm,width=10cm}
\caption[]{Fit of theoretical SED's computed for C-model
atmosphere 2000/5.0 (Tsuji 1999) to Kelu1 spectrum. $D$ factors
showed in the Fig. are used to simulate the {\em extra} depletion of several
species into grains.}
\end{figure}

\begin{figure}
\psfig{file=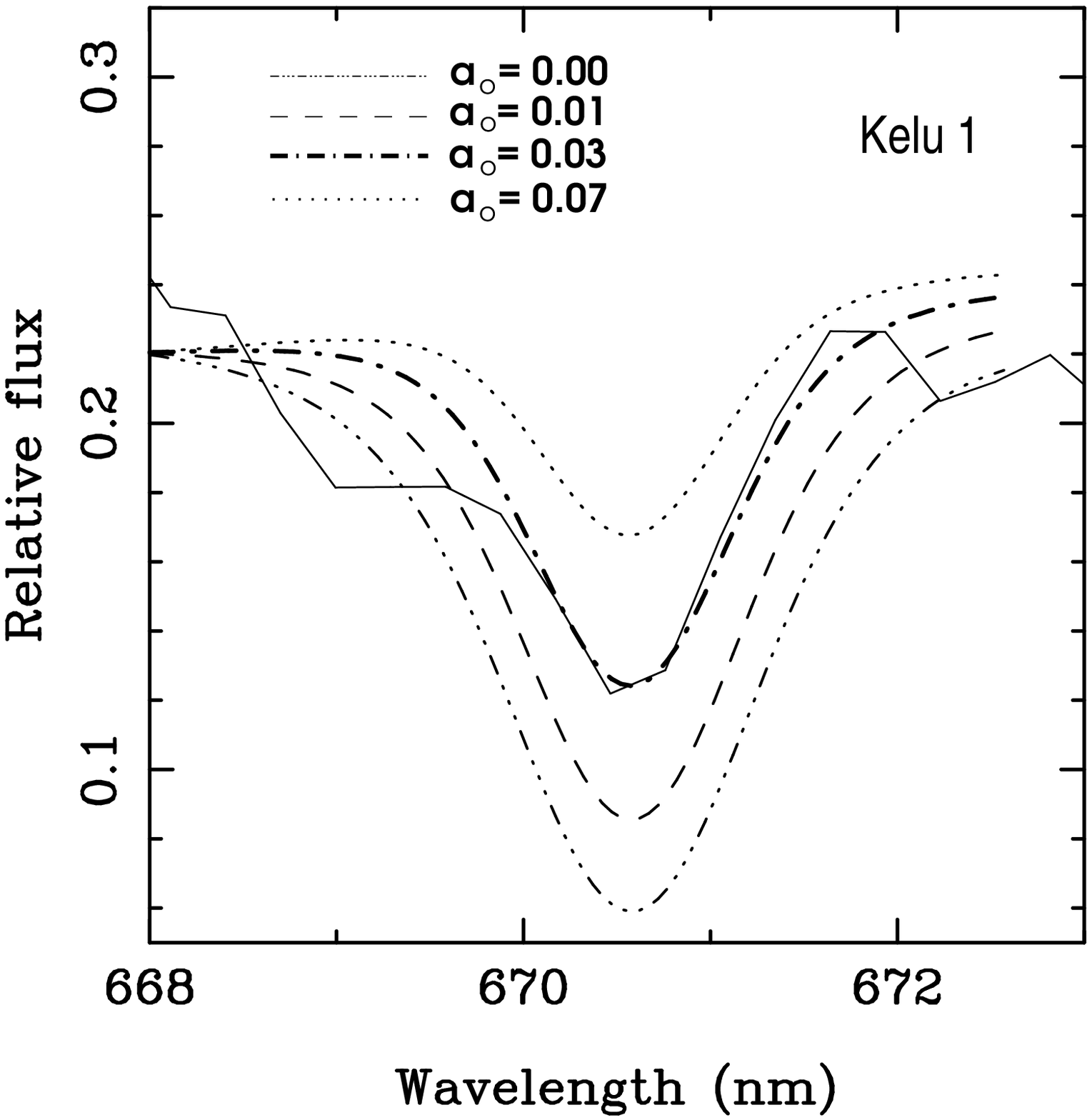,height=7cm,width=10cm}
\caption[]{Fit to Li I $\lambda$ 6708 nm
resonance doublet in Kelu1 spectrum. Computations were carried
out for log N (Li) = 3.0 and different parameters of the AdO}
\end{figure}

  Our computations show that  lithium lines are very
sensitive to the additional absorption (AdO) that we need to
incorporate in the  spectral synthesis if we want  to  explain
the  observed  broad  spectral energy distribution(Fig.4).
In  Table 1  we
give the predicted equivalent widths 
$W_{\lambda}$ of the Li\,{\sc i}  resonance  doublet  at
670.8\,nm  for  L-dwarf's    model    atmospheres
(2000--1000\,K) considered in this work. We found: \\

-- In the AdO-free case
(second column in  the  table),  we  would expect for the
``cosmic'' values of log N(Li) rather strong neutral Li resonance
lines in the spectra of objects as cool as DenisP\,J0205--1159
and Gl\,229B. \\

-- The chemical equilibrium of  Li-contained
species  still  allow  a  sufficient number of Li atoms to
produce a rather strong resonance  feature.

-- Our
computations indicate that L -dwarfs with moderate  dust
opacities  should  show  the
Li\,{\sc i} resonance doublet if they had preserved this  element
from nuclear burning, and consequently the lithium test can still
be applied.

-- Temporal variations of the dusty opacities may  originate some
kind of ``meteorological'' phenomena occurring in these cool
atmospheres. Lithium lines (as well other lines) may be severely
affected by the effect  (see Pavlenko et al. 2000 for more
details).

\begin{table}
\caption[]{Equivalent widths of  the  Li\,{\sc
i} resonance doublet at 670.8\,nm computed for the C-type Tsuji's
(1999)  model  atmospheres,  cosmic    Li    abundance
(log\,$N$(Li)\,=\,3.2)    and    gravity       log\,$g$\,=\,5.0.}
\begin{center}
\begin{tabular}{crrr}
\hline\hline
\multicolumn{1}{c}{}   &    \multicolumn{3}{c}{$a_{\circ}$}    \\
\multicolumn{1}{c}{$T_{\rm eff}$}                &
\multicolumn{3}{c}{\rule{2.5cm}{0.1mm}} \\ \multicolumn{1}{c}{} &
\multicolumn{1}{c}{0.00}    &    \multicolumn{1}{c}{0.01}       &
\multicolumn{1}{c}{0.10}    \\      \multicolumn{1}{c}{}        &
\multicolumn{3}{c}{\rule{2.5cm}{0.1mm}}                \\
\multicolumn{1}{c}{(K)} & \multicolumn{3}{c}{$W_{\lambda}$(\AA)} \\
\hline
  1000 & 17 &  8 & 0.6  \\
  1200 & 30 & 12 & 0.7  \\
  1400 & 42 & 21 & 0.9  \\
  1600 & 40 & 24 & 1.6  \\
  2000 & 23 & 16 & 3.6  \\
\hline
\end{tabular}
\end{center}
\end{table}

\section{Conclusions}

Finally, we have arrived to the following conclusions:

\begin{itemize}

\item Processes of formation of Li lines in L- and M-spectra
differ significantly:

-- for M-dwarfs the main problem to be solved is blending of
lithium lines by molecular lines,

-- in the case of L-dwarfs we deal with a menagerie of different
processes: depletion of lithium atoms into molecules and grains, "dusty
effects", meteorological phenomena, stratification effects, etc...

\item We can fit the optical spectra of L-dwarfs in the frame  of
 our simple model.

\item  Using our model we may perform a numerical analysis of the
   L-dwarf spectra (at least in the sense of the Li abundance determination).

\item The basic algorithm of the ``lithium test'' may be used
even for the assessment of the coolest L-dwarfs.

\end{itemize}

\section{Acknowledgements}

I thank IAU, LOC and SOC of the IAU Symposium N 198
for the financial support of my participation. I'm grateful to
R. Rebolo and M.R. Zapatero Ozorio
(IAC) for the fruitful collaboration and for providing the
observational data in electronic form; to T.Tsuji, F. Allard and
P.Hauschildt for providing model atmospheres in digital form.
Partial financial support was provided by the Spanish DGES
project no. PB95-1132-C02-01.

\end{document}